\documentclass[
 reprint,
 superscriptaddress,
 amsmath,amssymb,
 aps,
]{revtex4-2}

\usepackage{graphicx}% Include figure files
\usepackage{dcolumn}% Align table columns on decimal point
\usepackage{bm}% bold math
\usepackage[colorlinks=true, citecolor=Blue, linkcolor=Navy, urlcolor=RoyalBlue]{hyperref}% add hypertext capabilities
\usepackage{multirow} 
\newcolumntype{P}[1]{>{\centering\arraybackslash}p{#1}}
\newcolumntype{M}[1]{>{\centering\arraybackslash}m{#1}}

\usepackage{diagbox}
\usepackage[svgnames,table]{xcolor}

\usepackage{caption}
\usepackage{subcaption}
\usepackage{ragged2e}

\usepackage{braket}
\usepackage{diagbox}  

\DeclareCaptionJustification{justified}{\justifying}

\newcommand{\LG}{\mathrm{LG}}
\newcommand{\LGmode}[2]{\ensuremath{\LG_{#1,#2}}}

\newcommand{\LGzero}{\LGmode{0}{0}}
\newcommand{\LGtwo}{\LGmode{2}{2}}
\newcommand{\LGthree}{\LGmode{3}{3}}
\newcommand{\LGsix}{\LGmode{0}{6}}
\newcommand{\LGnine}{\LGmode{0}{9}}
\newcommand{\LGpl}{\LGmode{p}{\ell}}
\newcommand{\LGol}{\LGmode{0}{\ell}}

\begin{document}

%\preprint{APS/123-QED}

\title{Improving Beam Quality in Gravitational-Wave Interferometers Illuminated by Higher-Order Laguerre-Gaussian Modes}

\author{Liu Tao}
\email{liu.tao@apc.in2p3.fr}
\affiliation{Universit\'e Paris Cit\'e, CNRS, Astroparticule et Cosmologie, F-75013 Paris, France}

\author{Yuefan Guo}
\affiliation{Nikhef, Science Park 105, 1098 XG Amsterdam, The Netherlands}

\author{Alberto Gatto}
\affiliation{Dipartimento di Elettronica, Informazione e Bioingegneria, Politecnico di Milano, Milan 20133, Italy}

\author{Eleonora Capocasa}
\affiliation{Universit\'e Paris Cit\'e, CNRS, Astroparticule et Cosmologie, F-75013 Paris, France}

\author{Jérome Degallaix}
\affiliation{Universit\'e Claude Bernard Lyon 1, CNRS/IN2P3, IP2I Lyon, UMR 5822, LMA, UAR 2034, Villeurbanne, F-69100, France}

\author{Massimo Granata}
\affiliation{Universit\'e Claude Bernard Lyon 1, CNRS/IN2P3, IP2I Lyon, UMR 5822, LMA, UAR 2034, Villeurbanne, F-69100, France}

\author{Matteo Tacca}
\affiliation{Nikhef, Science Park 105, 1098 XG Amsterdam, The Netherlands}

\author{Matteo Barsuglia}
\affiliation{Universit\'e Paris Cit\'e, CNRS, Astroparticule et Cosmologie, F-75013 Paris, France}

\date{\today}

\begin{abstract}
Higher-order Laguerre-Gaussian (LG) laser modes have been proposed to reduce test-mass thermal noise in laser interferometric gravitational-wave detectors, owing to their more homogeneous intensity profiles compared to the currently employed fundamental Gaussian beam. However, LG beams such as the \LGthree{} mode suffer significant beam quality degradation in Fabry-Perot arm cavities in GW detectors with realistic state-of-the-art mirror surface figure errors, due to scattering into degenerate modes of the same order, which are resonantly enhanced by shared cavity resonance conditions. In this work, we investigate an alternative ``donut-shaped'' \LGol{}-like mode, specifically the \LGsix{} mode, and demonstrate strategies to improve its performance. These include the introduction of a tailored circular mirror mask with anti-reflective coating in the central region, which selectively increases the losses of parasitic degenerate modes while minimally impacting the \LGsix{} mode due to its limited overlap with the masked area. We further assess the marginal benefits of anticipated improvements in mirror surface figure errors and the potential reduction of cavity finesse. We demonstrate that these strategies can reduce the average contrast defect by more than two orders of magnitude and lower the mode loss by nearly a factor of five, achieving performance at or below the typical values observed in current detectors. This work opens up new research and development pathways for employing \LGol{}-type modes that achieve significant thermal noise reduction while maintaining beam quality and optical performance comparable to current gravitational-wave interferometers.

\end{abstract}

\maketitle

%\tableofcontents

\section{Introduction\label{sec-intro}}
\begin{figure}[t]
    \centering
    \includegraphics[width=\linewidth]{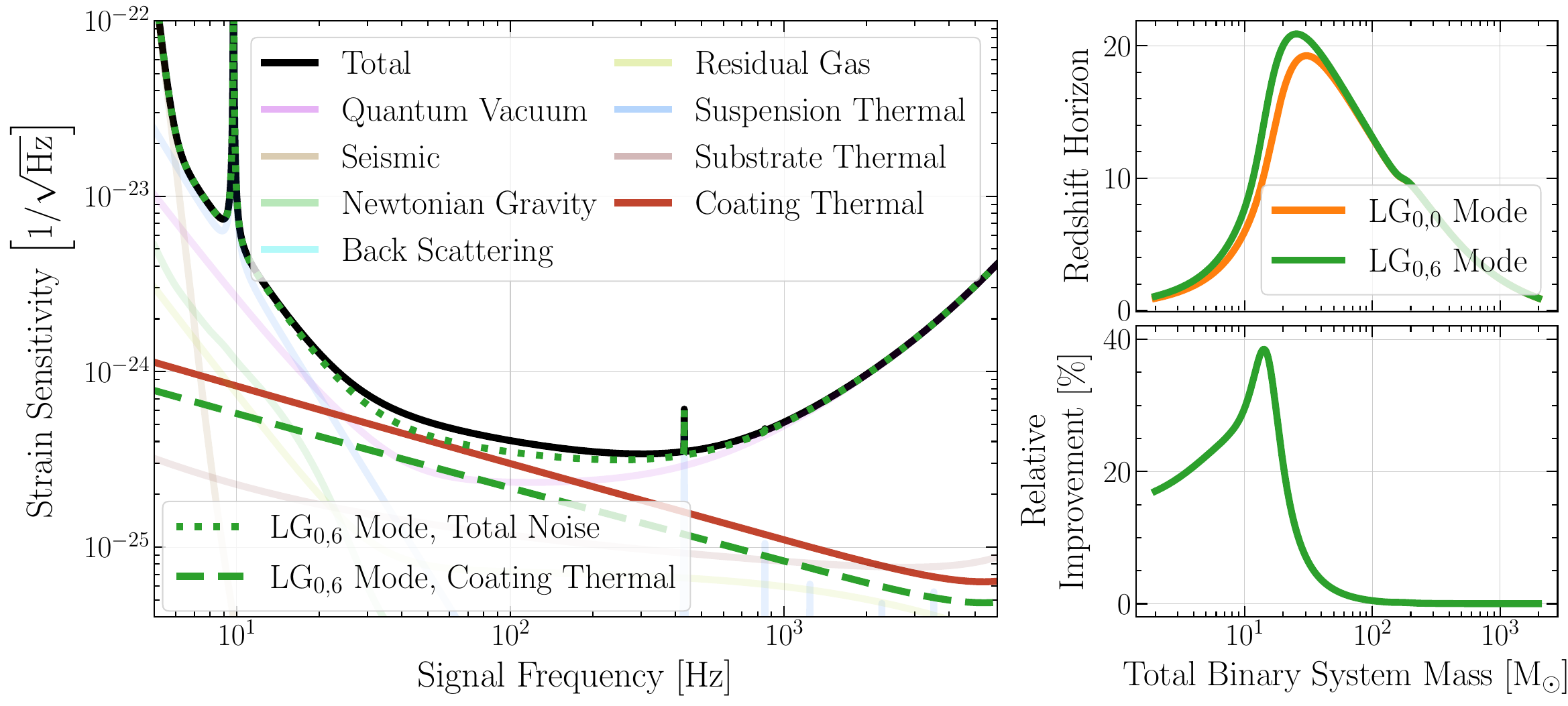}
    \caption{Noise budget for the Einstein Telescope high-frequency interferometer using the fundamental \LGzero{} and the higher-order \LGsix{} modes. \textit{Left:} Strain sensitivity, with total noise in black and individual contributions in color; coating thermal noise dominates near 100~Hz. The \LGsix{} case is shown with dashed (coating thermal noise) and dotted (total noise) lines. \textit{Right:} Cosmological redshift detection horizons for equal-mass, nonspinning binary black holes versus source-frame mass (top), and the relative improvement using \LGsix{} (bottom).}
    \label{fig-sensitivity}
\end{figure}

Test mass thermal noise, arising from the intrinsic Brownian motion of molecules in both the coating and bulk substrate of the interferometer mirrors, poses a major limitation to the sensitivity of current and next-generation gravitational-wave (GW) detectors~\cite{Capote_2025, Acernese_2015, CEHorizonStudy, ETDesignReportUpdate2020}. In the most sensitive frequency band around $100~\mathrm{Hz}$, thermal noise constitutes a dominant contribution to the overall noise budget, as illustrated in Fig.~\ref{fig-sensitivity} for the high-frequency interferometer of the Einstein Telescope~\cite{ETDesignReportUpdate2020}. 

Achieving the target sensitivity of future detectors, therefore, requires a substantial reduction of thermal noise. Contemporary mitigation strategies primarily focus on improvements in test-mass materials, mirror coatings, and operating conditions. These include the development of low-mechanical-loss coatings, as well as cryogenic operation using alternative substrates such as crystalline silicon or sapphire, both of which are candidate technologies for third-generation detectors such as the Einstein Telescope~\cite{mirrorcoating, PhysRevLett.127.071101, Cole_2023, Adhikari_2020}. These approaches aim to yield a reduction in coating thermal noise by approximately a factor of 2 compared to current detectors~\cite{PostO5Report:2022}.

A complementary approach to mitigating test mass thermal noise is to alter the distribution of laser power on the test masses. In current and planned detector designs, this can be achieved by adjusting the arm-cavity geometry, and therefore the cavity g-factor, to produce a larger spot size on the test masses while still using the fundamental Gaussian mode~\cite{ETDesignReportUpdate2020}. An alternative strategy, explored in this work, is to use alternative laser beam shapes, in particular beams with broader and more homogeneous intensity profiles than the Gaussian mode currently employed~\cite{Mours_2006, Vinet2009zz, PhysRevD.82.042003}. Such spatially extended beams enable more effective averaging of the random Brownian fluctuations of the test mass over the mirror surface, thereby reducing the resulting thermal noise~\cite{PhysRevD.82.042003}. 

\begin{figure*}[t]
    \centering
    \includegraphics[width=\linewidth]{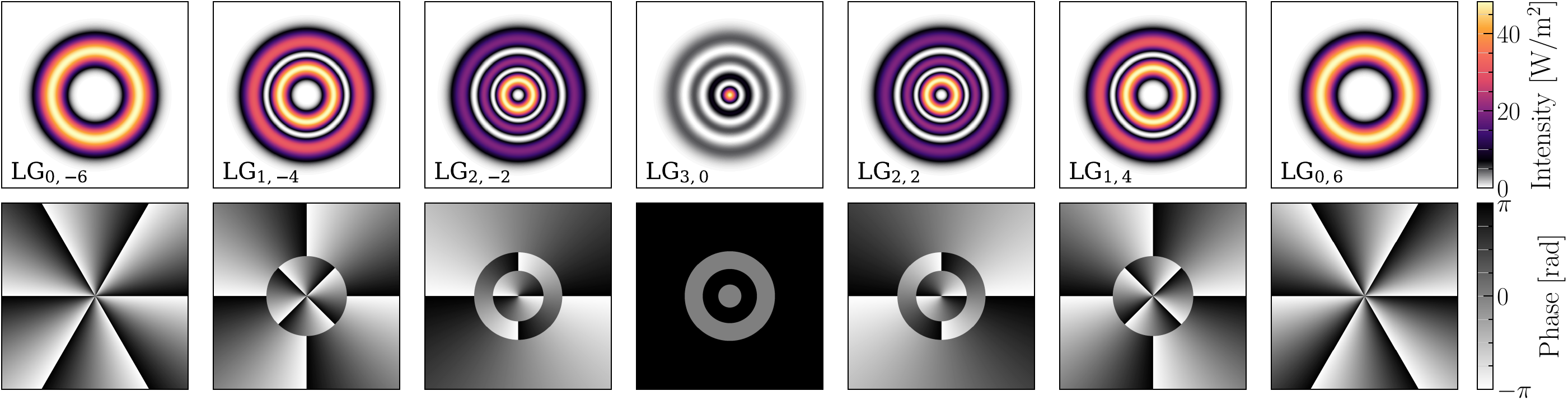}
    \caption{Intensity (top) and phase (bottom) profiles of all sixth-order Laguerre-Gaussian \LGpl{} modes.}
    \label{fig-intensity_phase_LGpl}
\end{figure*}

While exotic beam configurations, such as ``mesa'' beam~\cite{PhysRevD.67.102004} and ``conical'' beam~\cite{PhysRevD.78.082002}, have been investigated and may offer significant reductions in thermal noise, their implementation typically requires specially figured mirror surfaces to properly match the optical phase front and maintain stable confinement within the cavity. In contrast, a well-studied class of spatial beams that are more compatible with the instrumental infrastructures in current GW detectors, such as spherical cavity mirror geometry, is the higher-order Hermite-Gaussian (HG) and Laguerre-Gaussian (LG) modes. Among these, the $\mathrm{HG}_{3,3}$ and \LGthree{} modes are often identified as favorable compromises between thermal-noise reduction, beam-size scaling, and experimental feasibility~\cite{PhysRevD.82.012002, bond2011, PhysRevD.84.102001, Sorazu_2013, PhysRevLett.105.231102, PhysRevD.90.122011, PhysRevD.92.102002, Tao2020, PhysRevD.103.042008, PhysRevLett.132.101402}. Although HG modes provide a more modest reduction in thermal noise compared to LG modes of the same order~\cite{PhysRevD.82.042003}, they exhibit greater robustness against interferometer imperfections. In particular, controlled astigmatism can be introduced to lift the mode degeneracy, thanks to the symmetry between the tangential and sagittal components, thereby significantly mitigating scattering into modes of the same order~\cite{Tao2020, PhysRevD.103.042008}. For LG modes, such as the widely studied \LGthree{} mode, the coating thermal noise power spectral density (PSD) is expected to be reduced by approximately a factor of 7 for a fixed beam radius. When the beam radius is instead rescaled to maintain a constant clipping loss on a finite-sized mirror, the reduction factor becomes around 2.6 relative to the fundamental Gaussian mode~\cite{PhysRevD.82.042003}. 

Other than their potential application in GW detectors, higher-order Laguerre-Gaussian modes, particularly those with non-zero azimuthal index, carry orbital angular momentum (OAM), which has been widely exploited in quantum optics as a high-dimensional degree of freedom for information encoding and quantum entanglement~\cite{Krenn_2017, Fontaine_2019}. Specifically, helically phased \LGpl{} beams characterized by an azimuthal phase term $\exp(i \ell \phi)$ carry orbital angular momentum of $\ell \hbar$ per photon, where $\ell$ is the azimuthal index (also known as the topological charge), $\phi$ is the azimuthal angle, and $\hbar = h / 2\pi$ is the reduced Planck constant~\cite{PhysRevA.45.8185}. For instance, in the context of light-matter interaction, structured optical fields carrying OAM provide enhanced control over microscopic degrees of freedom, enabling optical tweezers capable of exerting torque on trapped particles and thereby allowing controlled rotation and micromechanical manipulation at the micro- and nanoscale~\cite{Porfirev}. Additionally, in optical communications, the orthogonality of LG modes enables multiplexing and demultiplexing of OAM channels, supporting parallel data transmission, enhanced spectral efficiency, and increased channel capacity~\cite{Wang:2012nkp}.

However, extensive studies, both in numerical simulations and tabletop experiments, have demonstrated that, despite their vast advantageous thermal-noise properties, LG modes are highly sensitive to realistic mirror-surface figure errors, especially astigmatism, arising from limitations in mirror polishing even in state-of-the-art optics, in Fabry-Perot arm cavities in GW detectors~\cite{PhysRevD.82.012002, bond2011, Sorazu_2013, PhysRevLett.105.231102, PhysRevD.90.122011, PhysRevD.92.102002}. This sensitivity arises primarily from scattering into co-resonant degenerate LG modes within the same mode order, which share identical Gouy phase accumulation and therefore satisfy the same resonance condition. This intra-order mode coupling can be systematically described by expanding mirror surface distortions in terms of Zernike polynomials $Z_n^m$. A non-zero coupling between an incident \LGpl{} mode and a scattered $\mathrm{LG}_{p',\ell'}$ mode requires matching of azimuthal indices such that $m = \Delta \ell = |\ell - \ell'|$, where $m$ is the azimuthal index of the corresponding Zernike term~\cite{bond2011}. In practice, the coupling is dominated by low-order aberrations, in particular the astigmatic terms $Z_2^{\pm 2}$, which drive the strongest mixing between modes with $\Delta \ell = 2$. As a result, such scattering is concentrated into the nearest degenerate neighbors in the mode subspace, most notably modes separated by $\Delta \ell = 2$ and, to a lesser extent, $\Delta \ell = 4$. Coupling to more distant modes is strongly suppressed due to the rapid decrease of higher spatial-frequency components of the mirror surface errors.

Limiting the mode coupling within this degenerate subspace and preserving beam purity are essential to maintaining the interferometer sensitivity. In this work, we investigate alternative ``donut-shaped'' \LGol{}-type modes such as the \LGsix{} and $\mathrm{LG_{0, -6}}$ modes shown in Fig.~\ref{fig-intensity_phase_LGpl}. A key feature of \LGol{}-like modes is the presence of a large central region with negligible intensity. This allows for tailored coating designs in which only the outer region of the test mass, where the beam intensity is concentrated, is coated with high-reflectivity (HR) coatings, while the central region is treated with an anti-reflection (AR) coating. Such a configuration introduces selective losses for unwanted degenerate modes, whose intensity profiles overlap significantly with the central region, while leaving the target mode largely unaffected. Furthermore, \LGol{}-like modes exhibit reduced scattering pathways and consequently lower losses compared to commonly studied modes such as \LGthree{} mode~\cite{bond2011}. For instance, in the presence of astigmatism, the dominant scattering from an injected \LGol{} mode occurs primarily into $\mathrm{LG}_{1,\ell-2}$, since the \LGol{} mode lies at the boundary of the degenerate mode subspace. This reduces the number of significant coupling channels by approximately a factor of two relative to \LGthree{}, which can scatter into multiple degenerate modes in both directions, such as $\mathrm{LG}_{2,5}$ and $\mathrm{LG}_{4,1}$ modes.

We evaluate several strategies to mitigate scattering into degenerate modes, including the use of a circular mirror mask to impose selective losses, reducing mirror surface figure errors, and the effect of lowering the cavity finesse. We consider a reduction in the mirror surface error from the state-of-the-art root-mean-square (RMS) level of 0.3~nm to 0.1~nm, as anticipated from near-future improvements in large mirror polishing techniques and metrology systems~\cite{Pinard:17, Billingsley2017_LIGO_P1700029}. We use a current Virgo-like configuration as a representative case study, as it provides experimentally measured mirror surface maps and enables direct comparison with measured optical performance, offering a realistic framework for evaluating the robustness of higher-order LG modes~\cite{Acernese_2015}. Combined with the use of a circular mask of variable radius and a reduction of the arm cavity finesse by a factor of two in a Virgo-like configuration, these approaches provide a viable pathway to improving beam quality and maintaining acceptable optical performance. We find, for example, that the average contrast defect can be reduced by more than two orders of magnitude, from $49994~\mathrm{ppm}$ to $352~\mathrm{ppm}$, by employing a circular mask with an optimal radius of $3.2~\mathrm{cm}$ when balancing the increased mode loss, assuming a 0.1~nm RMS surface error and a reduced cavity finesse by half. This value is within a factor of two of the typical contrast defect measured in Virgo~\cite{VirgoCollaboration:25, virgo_logbook_68368}. A larger circular mask of $4~\mathrm{cm}$ yields the minimum achievable contrast defect, reaching around $140~\mathrm{ppm}$ for the \LGsix{} mode, which is approximately a factor of two below typical Virgo measurements. However, this configuration incurs higher mode loss and therefore represents a trade-off between minimizing contrast defect and maintaining acceptable optical losses. Under the same set of assumptions, the mode loss is also reduced significantly, from $9.82\%$ to $2.05\%$, which is slightly below the estimated cavity loss in Virgo of $2.12\%$~\cite{VirgoCollaboration:25}. 

These results demonstrate that, by combining circular masking with engineered selective losses for degenerate modes, improved mirror surface quality, and reduced cavity finesse, the \LGsix{} mode can maintain beam quality and optical performance at levels comparable to current gravitational-wave detector operation. This study serves as a representative case for \LGol{}-type modes more broadly. Comparable improvements in beam quality have also been observed for other \LGol{} modes; for instance, the ninth-order $\mathrm{LG}_{0,9}$ mode exhibits similar enhancement in modal purity when the same strategies are applied. In addition, improved modal purity in such higher-order LG modes in high-precision metrology systems can further enhance their effectiveness in other optics applications that rely on well-defined OAM states, including high-dimensional quantum information processing, optical manipulation, and OAM-resolved spectroscopy. In particular, as a result of improved modal purity of the information-carrying LG modes, reduced mode mixing and inter-modal cross-talk directly translate to higher fidelity, efficiency, and measurement precision in these metrological systems~\cite{yan2014high, Liu:26}.

The paper is organized as follows. In \S\ref{sec-background}, we introduce Laguerre-Gaussian modes and their associated coating thermal noise reduction factors. In \S\ref{sec-strategy}, we present the proposed strategies for mitigating beam quality degradation of \LGol{}-type modes in the presence of realistic mirror surface figure errors, including the use of a circular mask, reduction of the surface RMS error, and reduction of the cavity finesse. In \S\ref{sec-simulation}, we describe the optical simulations, as well as the resulting improvements in the figures of merit used to evaluate performance via Monte Carlo analysis. Finally, in \S\ref{sec-conclusion}, we conclude with a discussion of the results, their implications, and potential future extensions.

\section{Background \label{sec-background}}
\subsection{Laguerre-Gaussian modes}
Laguerre-Gaussian modes form a complete orthonormal set of solutions to the paraxial Helmholtz equation in cylindrical coordinates. They accurately describe free-space Gaussian beam propagation and provide a good approximation to the eigenmodes of optical resonators with spherical mirrors under circular symmetry. Each LG mode (\LGpl{}) is characterized by two integers: the radial index $p \geq 0$, which determines the number of radial nodes, and the azimuthal index $\ell$, which specifies the helical phase structure and corresponds to the number of $2\pi$ phase windings in the azimuthal direction~\cite{Bond2017}. The field distribution of an \LGpl{} mode can be written as
\begin{equation}
\begin{aligned}
\psi_{p,\ell}(r,\phi,z) =\;& \frac{1}{w(z)} \sqrt{\frac{2 p!}{\pi (p+|\ell|)!}}
\left( \frac{\sqrt{2}\, r}{w(z)} \right)^{|\ell|} 
L_p^{|\ell|}\!\left( \frac{2 r^2}{w^2(z)} \right) \\
& \times \exp\!\big[i(2p+|\ell|+1)\,\Psi(z)\big] \\
& \times \exp\!\left(-\frac{r^2}{w^2(z)} - i \frac{k r^2}{2 R_c(z)} + i \ell \phi \right),
\end{aligned}
\end{equation}
where $k$ is the wave number, $\Psi(z)$ is the Gouy phase, $L_p^{|\ell|}(x)$ denotes the generalized Laguerre polynomial, and $w(z)$ and $R_c(z)$ are the beam radius and radius of curvature at longitudinal position $z$, respectively.

\begin{figure}[t]
    \centering
    \includegraphics[width=\linewidth]{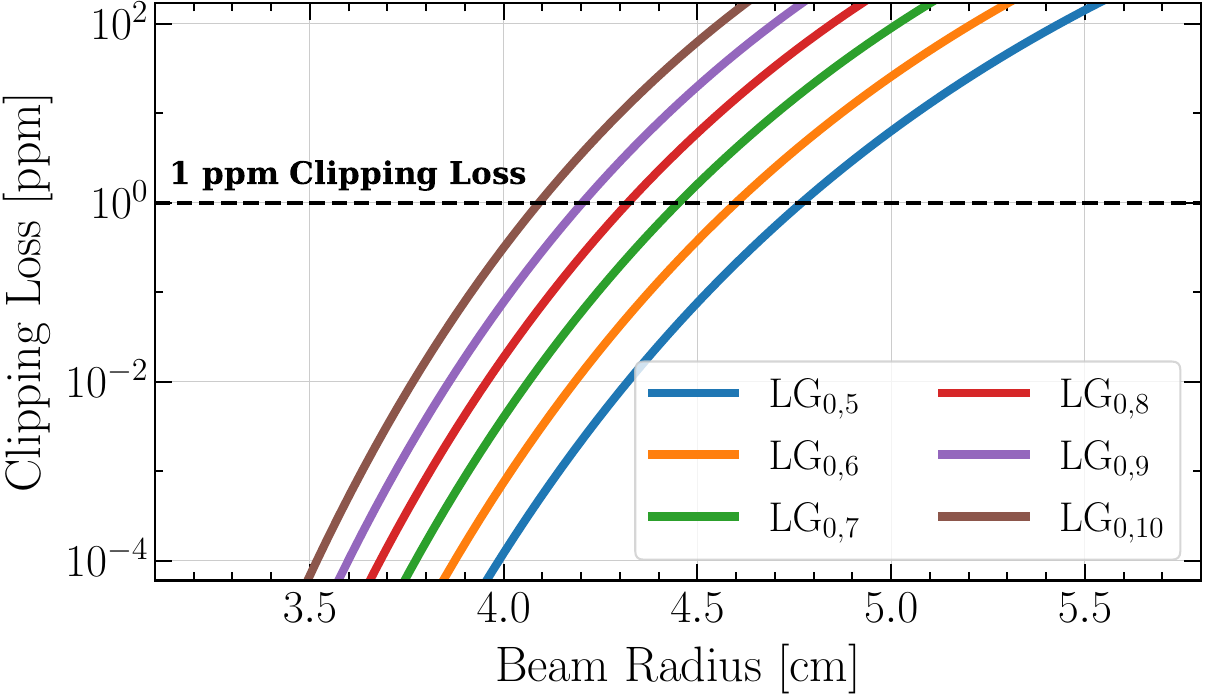}
    \caption{Clipping loss as a function of beam radius for \LGol{} modes with $\ell=5$--$10$ on a Virgo-scale mirror of radius $R=0.17\,\mathrm{m}$. The horizontal dashed line indicates the 1~ppm clipping loss requirement. }
    \label{fig-clipping_loss_LG0l}
\end{figure}

\begin{figure}[t]
    \centering
    \includegraphics[width=\linewidth]{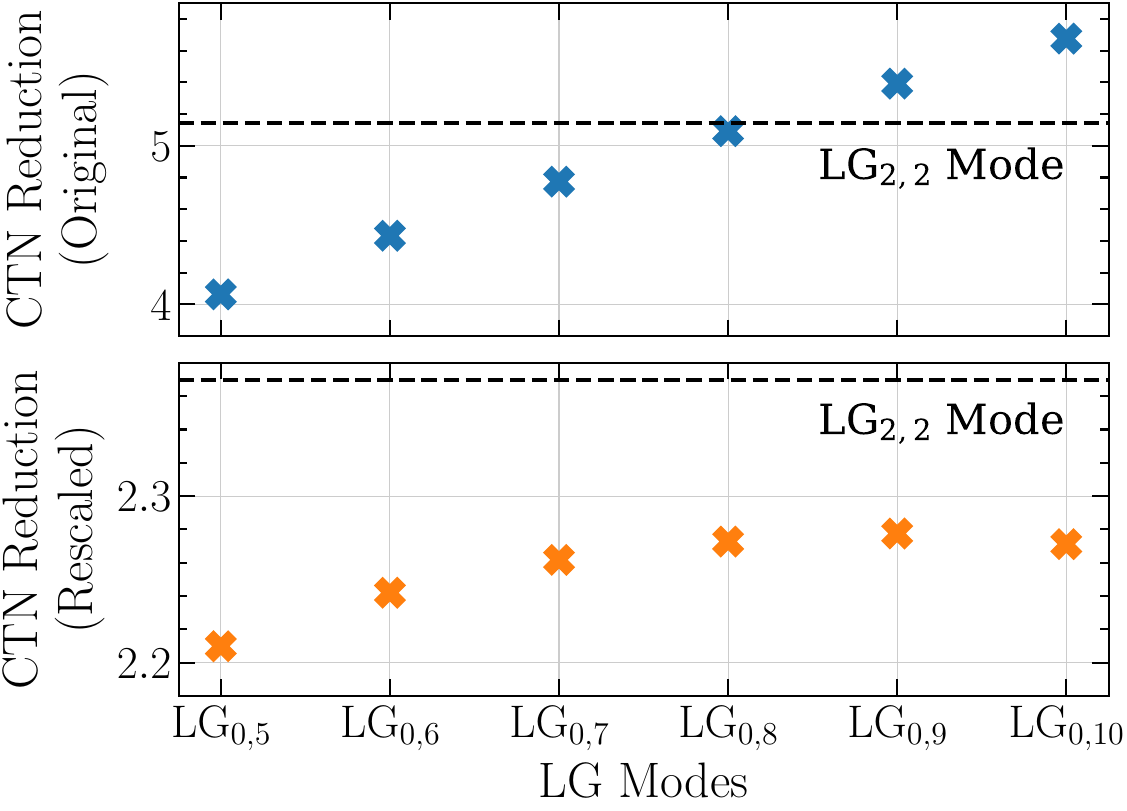}
    \caption{Coating thermal noise (CTN) power spectral density reduction factors for the ``donut-shaped'' \LGol{} modes with $\ell=5$--$10$ relative to the fundamental mode. The top panel shows the reduction factor assuming the same beam radius for all modes, while the bottom panel shows the reduction factor when beam radii are rescaled to maintain the same clipping loss. The horizontal dashed line indicates the reduction factor for the commonly studied sixth-order \LGtwo{} mode for comparison.}
    \label{fig-thermal_noise_reduction_LG0l}
\end{figure}

The mode order of an LG mode is defined as $N = 2p + |\ell|$. All modes of the same order $N$ accumulate the same Gouy phase and are therefore degenerate in an ideal spherical-mirror cavity. The number of modes within a given order is $N + 1 = 2p + |\ell| + 1$. Fig.~\ref{fig-intensity_phase_LGpl} illustrates the intensity (top) and phase (bottom) distributions of representative \LGpl{} modes with order $N = 6$.

\subsection{Thermal noise reduction}
Owing to their more homogeneous intensity distribution, these modes reduce the contribution of thermal noise such as the coating thermal noise (CTN), whose power spectral density scales with the integral of the squared beam intensity over the mirror surface (the $\mathcal{L}^2$ norm). For a given \LGpl{} mode, the reduction factor can be expressed as~\cite{PhysRevD.82.042003}
\begin{equation}
\mathcal{J}_{p,l} = 2 \int_0^{\infty} e^{-2x} L_p(x)^2 \, L_{p+l}(x)^2 \, dx \, ,
\label{equ-jpl}
\end{equation}
where $L_p(x)$ denotes the Laguerre polynomial of order $p$. 

The coating thermal noise PSD also scales inversely with the square of the beam radius~\cite{PhysRevD.82.042003}. For higher-order \LGpl{} modes with broader intensity distributions, maintaining a fixed level of power loss at the edges of a finite-sized mirror (i.e., the clipping loss) requires reducing the beam radius. The clipping loss is computed as
\begin{equation}
L_{p,l}(\omega) = 1 - \int_{0}^{2\pi} d\phi \int_{0}^{R} dr \, r \, I_{p,l}(\omega),
\end{equation}
where $I_{p,l}(\omega)$ denotes the normalized intensity profile of the \LGpl{} mode on a mirror of radius $R$, as illustrated by the example intensity distributions in Fig.~\ref{fig-intensity_phase_LGpl}. Fig.~\ref{fig-clipping_loss_LG0l} shows the resulting clipping loss as a function of beam radius for a Virgo-scale mirror with $R=0.17\,\mathrm{m}$, considering \LGol{}-type modes with $\ell=5$--$10$. The horizontal line indicates the 1~ppm clipping loss requirement. For instance, the maximum allowable beam radius for the \LGsix{} mode is approximately $4.6\,\mathrm{cm}$ to maintain a clipping loss below 1~ppm. As the azimuthal index $\ell$ of \LGol{} modes increases, the mode becomes increasingly spatially extended. Consequently, progressively smaller beam radii are required to satisfy the same clipping loss constraint. This beam-size rescaling partially balances out the thermal noise reduction advantage offered by wider beams. Consequently, there exists a trade-off between improved spatial averaging of thermal fluctuations, thus lower thermal noise, and the constraint imposed by clipping loss.

Fig.~\ref{fig-thermal_noise_reduction_LG0l} shows the coating thermal noise reduction factor for \LGol{}-like modes, obtained by evaluating Eq.~(\ref{equ-jpl}) numerically. The top panel presents the reduction factor for modes with a fixed beam radius, while the bottom panel shows the reduction factor when the beam radii are rescaled such that all modes exhibit the same clipping loss of $1~\mathrm{ppm}$, as indicated in Fig.~\ref{fig-clipping_loss_LG0l}. For reference, the reduction factor for the commonly studied \LGtwo{} mode is indicated by the horizontal dashed line.

In the beam-size-rescaled case, the thermal noise reduction for \LGol{}-like modes initially improves with increasing azimuthal index $\ell$, reflecting their broader intensity profiles. However, as $\ell$ increases further, the required beam radius reduction to maintain constant clipping loss becomes more significant, eventually limiting the achievable noise reduction. The optimal performance is reached at the \LGnine{} mode. Notably, the coating thermal noise reduction factor by \LGsix{} mode is 2.24, which is only $1.6\%$ lower than that of the ninth-order \LGnine{} mode, while offering advantages such as a smaller degenerate mode set, thus fewer scattering loss channels, and potentially higher generation efficiency. Furthermore, the reduction in thermal noise from the \LGsix{} mode to the \LGtwo{} mode (both of the same order) is only approximately $5.4\%$, indicating that \LGol{}-like modes retain significant advantages for thermal noise mitigation.

The implementation of higher-order modes is expected to be most relevant for next-generation gravitational-wave detectors operating in the high-power, thermal-noise-limited regime~\cite{HOM_thermal}. Fig.~\ref{fig-sensitivity} compares the total strain sensitivity of the Einstein Telescope high-frequency interferometer if the currently used fundamental Gaussian beam is replaced by the higher-order \LGsix{} mode (green); the coating thermal noise and the corresponding total strain noise are shown as dashed and dotted lines~\cite{2020ascl.soft07020R}. All technical noise sources except coating thermal noise are displayed in the background with reduced opacity to better highlight the dominant thermal noise. The reduction in coating thermal noise amplitude spectral density for the \LGsix{} mode, $\sqrt{2.24}\approx 1.50$, is illustrated in the figure. The right panels of Fig.~\ref{fig-sensitivity} present the resulting cosmological redshift detection horizons for nonspinning, equal-mass binary black hole mergers as a function of the total source-frame mass for the fundamental and the higher-order mode (top), together with the relative improvement with respect to the fundamental Gaussian mode (bottom). As indicated, the use of \LGsix{} mode can yield close to 40\% improvement in the sky-averaged detection range for binary mergers due to the associated reduction in coating thermal noise.

\section{Beam quality improvement strategies \label{sec-strategy}}
\subsection{Selective loss with mirror mask }

\begin{figure}[t]
    \centering
    \includegraphics[width=\linewidth]{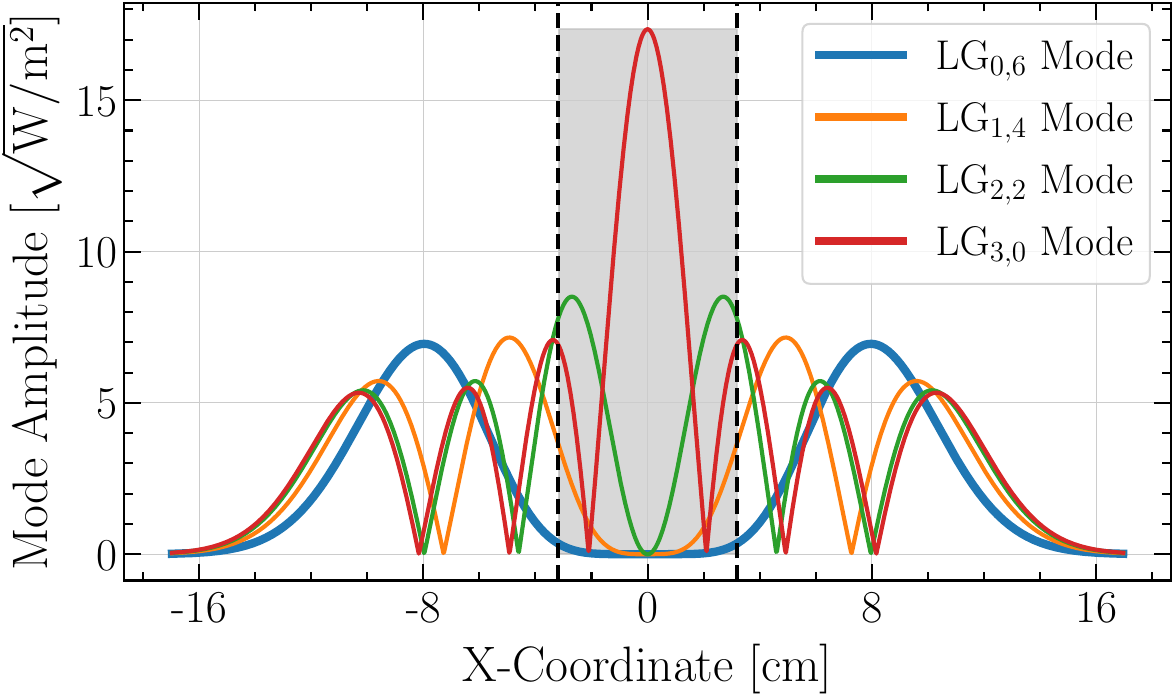}
    \caption{Cross-sectional profiles of the mode amplitude for sixth-order \LGpl{} modes. The central circular mask region with AR coating is indicated by the gray area between the two vertical dashed lines. The \LGsix{} mode (blue) has negligible overlap with the AR-coated region, whereas the degenerate modes exhibit significant overlap.}
    \label{fig-amplitude_LGpl_1D_gray_area}
\end{figure}

A technique proposed to improve the modal purity of higher-order Laguerre-Gaussian modes is the introduction of spatially selective losses, e.g.\ via absorbing or anti-reflection coatings placed at intensity nodes, to lift degeneracy and suppress unwanted modes \cite{PhysRevD.84.102001, Yamamoto2011_LIGO_T1100220}. While this approach was historically considered for modes such as the \LGthree{} mode, its practical implementation is challenging due to the complex radial structure: placing lossy elements between intensity rings leads to excess scattering, since the field derivative is large and finite-width features inevitably overlap regions of non-zero intensity. 

Motivated by this limitation, the alternative ``donut-shaped'' \LGol{} modes provide a more favorable configuration. Owing to their central phase singularity, these modes have a naturally dark central region in which both the optical intensity and its spatial gradient are small~\cite{LG09_2014}. This feature makes it possible to modify the coating near the center of the mirror while introducing only a limited perturbation to the target \LGol{} mode. Thus, we consider a circularly symmetric coating design in which the HR coating is applied only to the outer region of the mirror, while the central circular region is either left uncoated (where the bare substrate
provides approximately 4\% reflectivity) or treated with an AR coating. In practice, depositing a coating onto only a portion of a mirror surface is typically achieved by inserting a mask between the substrate and the ion beam sputter source; however, during repeated sputtering, the material deposited near the inner edge of the mask projection can vary in thickness, creating a 2~mm radial transition
zone on the mirror. Regardless, our first simulations showed that depositing an AR-coating on the central region and the detailed shape of the boundary between the AR- and HR-coated regions do not significantly affect the main results: an uncoated central region gives slightly lower loss but a larger contrast defect than the AR-coated case, while a linear transition gives lower contrast defect but higher loss, with differences at the level of $\sim 1\%$ and $\sim 10\%$, respectively. Thus, a simplified model featuring a sharp boundary between the outer HR coating and the central AR-coated region, without any transition zone, was used for the remainder of the simulations.

Such a tailored coating profile introduces significant optical loss for unwanted degenerate modes of the same order, thereby reducing their effective finesse, while having minimal impact on the injected \LGol{} mode. This selective loss mechanism mitigates parasitic scattering into degenerate modes, improving modal purity and, consequently, the interferometer sensitivity. Moreover, the size of this central region increases with the azimuthal index $\ell$, making such approaches increasingly practical for higher-order modes. In this work, the AR-coated mask is applied only to the end mirror, since introducing masks on both the input and end mirrors significantly increases the cavity loss while providing only limited improvement in mode purity~\cite{Guo2025Thesis}. A complete assessment of applying a corrective mask to the input mirror will require future studies in a full dual-recycled interferometer configuration to evaluate the impact of leaked higher-order modes into the recycling cavities and their implications for interferometer sensing and control.

\begin{figure}[t]
    \centering
    \includegraphics[width=\linewidth]{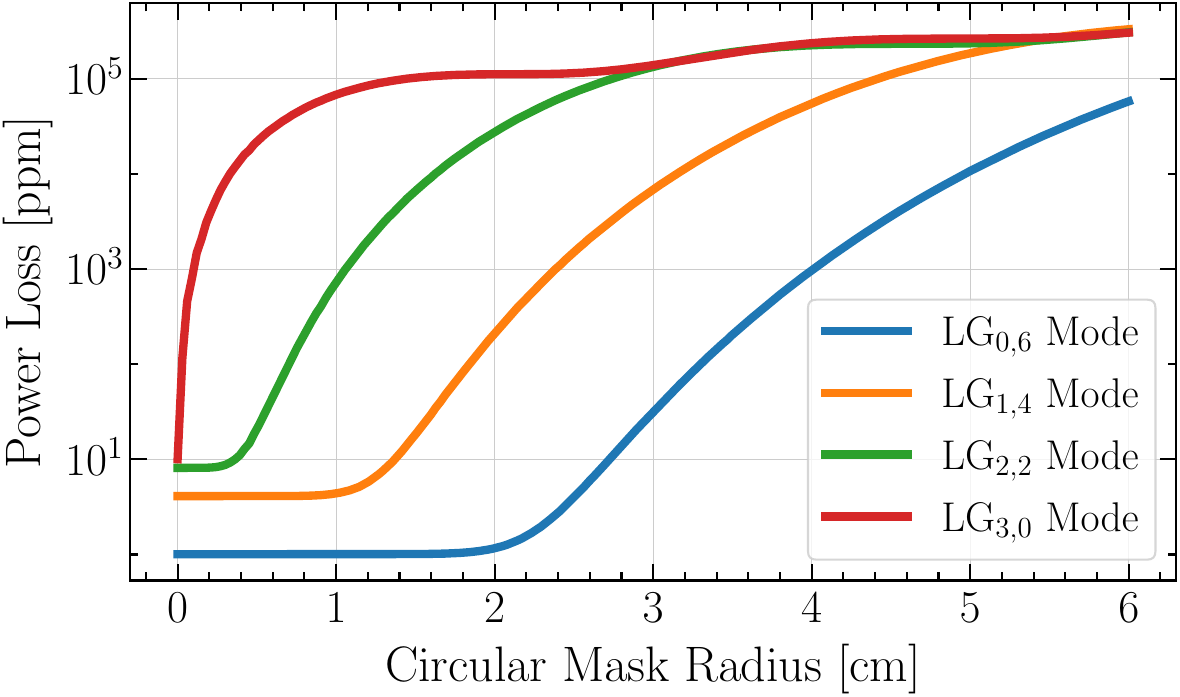}
    \caption{Power loss due to the circular mask and aperture for sixth-order \LGpl{} modes as the mask radius increases. A tailored circular mask can introduce significant loss for the degenerate modes while maintaining low loss for the \LGsix{} mode. The initial loss with no circular mask arises from beam clipping at the mirror edge.}
    \label{fig-power_loss_LGpl_r_mask}
\end{figure}

Fig.~\ref{fig-amplitude_LGpl_1D_gray_area} shows the amplitude profiles of LG modes of order six. All modes are assumed to have the same beam radius of $4.6~\mathrm{cm}$ on a Virgo-like circular test mass with radius $R = 0.17~\mathrm{m}$, corresponding to a clipping loss of $1~\mathrm{ppm}$ for the \LGsix{} mode. The \LGsix{} mode exhibits its peak amplitude around $0.08~\mathrm{m}$, while maintaining a near-zero amplitude in the central region. Introducing a circular AR-coated mask, indicated by the shaded region between the vertical dashed lines, results in substantial additional loss for degenerate modes such as $\mathrm{LG}_{3,0}$, whose amplitude overlaps significantly with the masked region.

Fig.~\ref{fig-power_loss_LGpl_r_mask} shows the power loss due to the circular mask and aperture as a function of mask radius for all degenerate modes of order six. As the mask radius increases, the power loss for the \LGsix{} mode remains approximately constant at the $1~\mathrm{ppm}$ clipping level up to a radius of around $2~\mathrm{cm}$. In contrast, other degenerate modes exhibit a much stronger dependence on the mask size: the $\mathrm{LG}_{3,0}$ mode experiences increased loss immediately, the $\mathrm{LG}_{2,2}$ mode shows a rise at $0.4~\mathrm{cm}$, and the $\mathrm{LG}_{1,4}$ mode at $1~\mathrm{cm}$. This behavior enables the design of an optimal mask radius that maximizes loss for parasitic modes while minimizing additional loss for the injected \LGsix{}. In the following, we quantify this trade-off by performing a statistical analysis over many randomized realizations of realistic mirror surface distortions. We use the averaged contrast defect and mode loss as performance metrics to identify the optimal mask size.

\subsection{Near-future reduction in mirror surface error}
Test masses in gravitational-wave detectors do not exhibit perfectly spherical surfaces due to limitations in current mirror polishing techniques and metrology systems. In particular, even state-of-the-art mirrors used in current detectors exhibit surface figure errors at the level of $0.3~\mathrm{nm}$ RMS. Such imperfections induce scattering from the injected \LGsix{} mode into other degenerate LG modes, which are also resonantly enhanced in optical cavities. This results in increased scattering losses and a degradation of the interferometer contrast at the dark port, arising from imbalanced modal content between the two arms at the central beamsplitter.

\begin{figure}[t]
    \centering
    \includegraphics[width=\linewidth]{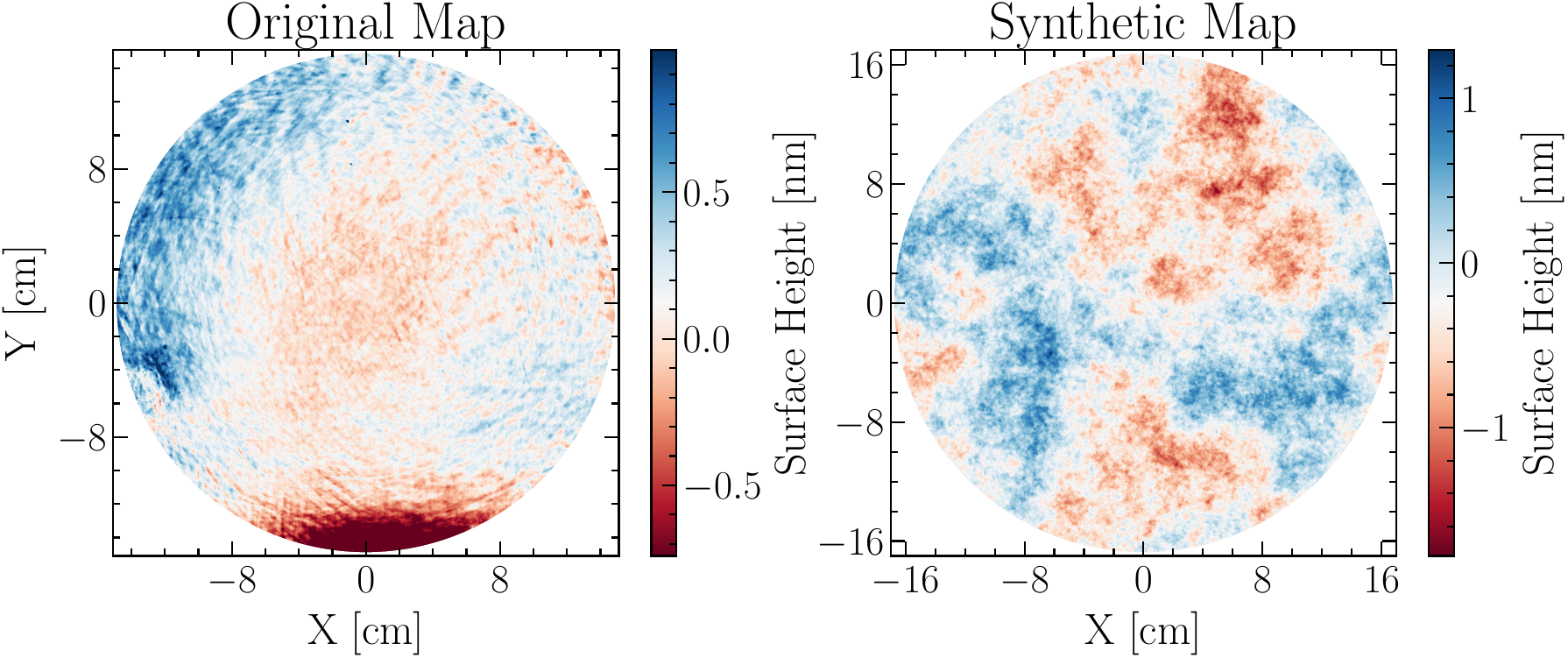}
    \caption{Measured mirror surface figure error for ITM04 in LIGO (left) and an example random synthetic map (right) that preserves the same 1D spatial spectrum as the original map. Both maps have the same integrated RMS surface error of 0.3~nm.}
    \label{fig-random_map_example}
\end{figure}

We simulate the impact of mirror surface errors using the widely adopted interferometer simulation software \texttt{Finesse}~\cite{brown_2025_12662017}, modeling the surface deformation as two-dimensional mirror maps that describe the height variation $\delta h(x,y)$ across the optic. Upon reflection, these surface distortions impart a spatially varying phase shift to the incident optical field, given by
\begin{equation}
\Phi(x,y) = \frac{4\pi}{\lambda}\,\delta h(x,y),
\label{equ-phasemap}
\end{equation}
where $\lambda$ is the laser wavelength. The left panel of Fig.~\ref{fig-random_map_example} shows an example of a measured uncoated mirror substrate in Advanced LIGO, specifically the ITM04 optic.

To assess the robustness of the \LGsix{} mode against such realistic surface perturbations, we perform Monte Carlo simulations over a large ensemble of randomized mirror maps. Each trial simulation generates randomized mirror surface maps, and the corresponding optical response of the cavity is evaluated.

\begin{figure}[t]
    \centering
    \includegraphics[width=\linewidth]{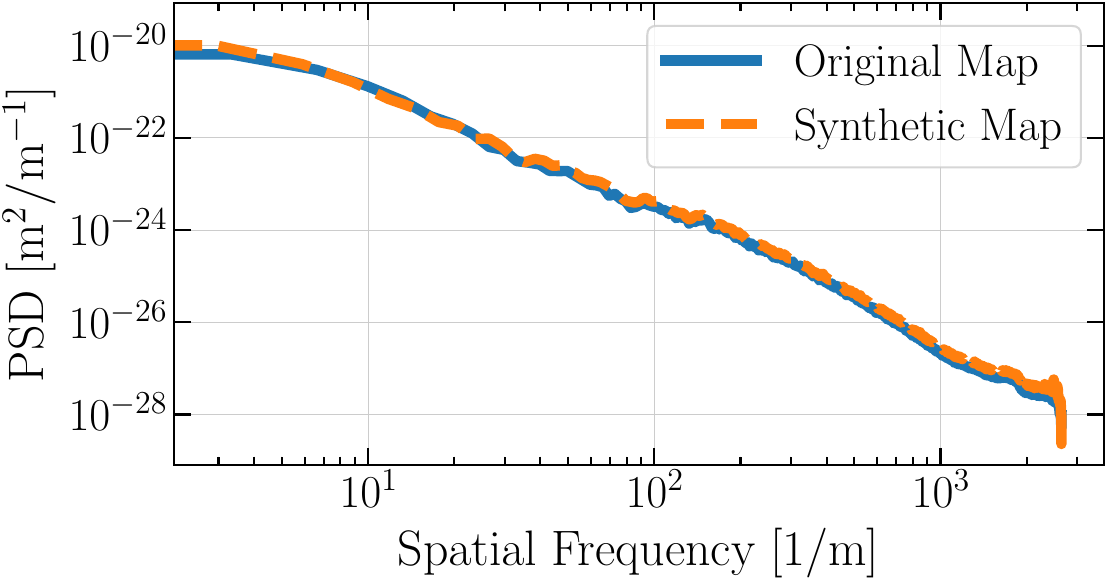}
    \caption{One-dimensional spatial Fourier power spectral density of the original map and the example synthetic map in Fig.~\ref{fig-random_map_example}, showing that both maps exhibit the same 1D spatial spectra.}
    \label{fig-random_map_radial_spatial_spectrum_1d}
\end{figure}

Synthetic mirror maps are generated to reproduce the spatial-frequency characteristics of the measured reference map. We first compute the two-dimensional Fourier transform of the measured surface map and obtain the corresponding two-dimensional power spectral density, which is then reduced to a one-dimensional radial profile by averaging over annular bins of constant radial spatial frequency. This one-dimensional radial amplitude spectrum is then interpolated back onto a two-dimensional spatial-frequency grid. Random phases, uniformly distributed between $0$ and $2\pi$, are assigned independently to each point of this two-dimensional frequency-domain map. The randomized complex spectrum is then transformed back to real space using an inverse FFT to obtain a synthetic surface map. An example of such a synthetic map is shown in the right panel of Fig.~\ref{fig-random_map_example}. Fig.~\ref{fig-random_map_radial_spatial_spectrum_1d} compares the one-dimensional PSDs of the measured and synthetic maps, demonstrating that they share the same spectral characteristics.

The RMS surface error of a mirror map can be computed from its PSD as
\begin{equation}
\sigma_{\mathrm{RMS}} = \sqrt{\int_{f_{\min}}^{f_{\max}} \mathrm{PSD}(f)\, df} \, .
\end{equation}

All generated maps are rescaled to yield the same spatial Fourier spectrum and integrated RMS. This work considers the improvement in beam quality associated with a reduction in the RMS error from $0.3~\mathrm{nm}$ to $0.1~\mathrm{nm}$.

\subsection{Reducing the cavity finesse}
In Fabry-Perot arm cavities in GW detectors, the degeneracy of higher-order spatial modes leads to their co-resonant enhancement, which amplifies the coupling of mirror surface imperfections into unwanted modes and exacerbates losses. This effect degrades beam quality and increases contrast defect at the interferometer output. Lowering the arm cavity finesse reduces the strength of this degeneracy-driven enhancement, thereby mitigating mode scattering and improving the robustness of higher-order mode operation. Thus, reducing the arm cavity finesse provides a viable solution to mitigate optical losses and contrast defects in gravitational-wave detectors. The resulting reduction in arm-cavity power buildup can be compensated by increasing the power recycling gain (PRG) through appropriate tuning of the power-recycling cavity, thereby restoring the circulating power while maintaining a broader optical bandwidth. For instance, in the Virgo configuration, as the arm cavity finesse is reduced by a factor of two, the power-recycling mirror transmission can be reduced by a factor of around 2.18 (from the nominal 4.835\% to 2.222\%), thereby increasing the power-recycling gain by a factor of two and maintaining a fixed total power gain.

However, lowering the arm cavity finesse also increases the interferometer bandwidth. To recover the original detector frequency response, the transmission of the signal recycling mirror must be retuned accordingly. For the Virgo configuration, in addition to the reduction in power recycling mirror transmission described above, increasing the signal recycling mirror transmission by a factor of $1.62$ (from 40.0\% to 64.8\%) is sufficient to restore the quantum-noise-limited sensitivity of the detector.

Although increasing the PRG leads to higher power buildup within the recycling cavity and raises concerns regarding thermal loading in the central interferometer, this is not expected to pose a significant limitation in the context of higher-order mode operation. Recent studies indicate that higher-order spatial modes exhibit a more favorable thermo-optical performance than the sharply-peaked fundamental mode, with reduced peak thermal gradients and associated scattering loss, owing to their wider intensity distributions~\cite{HOM_thermal}. As a result, the increased thermal load in the recycling cavity can be potentially mitigated, and the use of elevated PRG in conjunction with lower arm cavity finesse is expected to remain compatible with stable operation, while offering improved robustness against realistic mirror surface errors. Nevertheless, a full dual-recycled interferometer model will be needed to assess whether the required recycling-cavity retuning introduces additional higher-order mode buildup, sensing and control challenges, or other system-level limitations.

\section{Optical performance evaluation with Monte Carlo simulation \label{sec-simulation}}

\subsection{Optical Simulation}

\begin{table}[b]
\centering
\caption{Beam radius and radius of curvature for a mirror with $R=0.17$~m in a Virgo-like symmetric cavity of length 2998~m, with clipping loss maintained at 1~ppm.}
\setlength{\tabcolsep}{10pt}
\begin{tabular}{c|c|c}
\hline 
\hline 
Mode & \LGzero{} & \LGsix{}  \\
\hline  
Beam Radius w [cm] & 6.47 & 4.60  \\ \hline 
Mirror RoC [m] & 1521.8 & 1596.9  \\ \hline \hline 
\end{tabular}
\label{tab-roc_homs}
\end{table}

\begin{figure}[t]
    \centering
    \includegraphics[width=\linewidth]{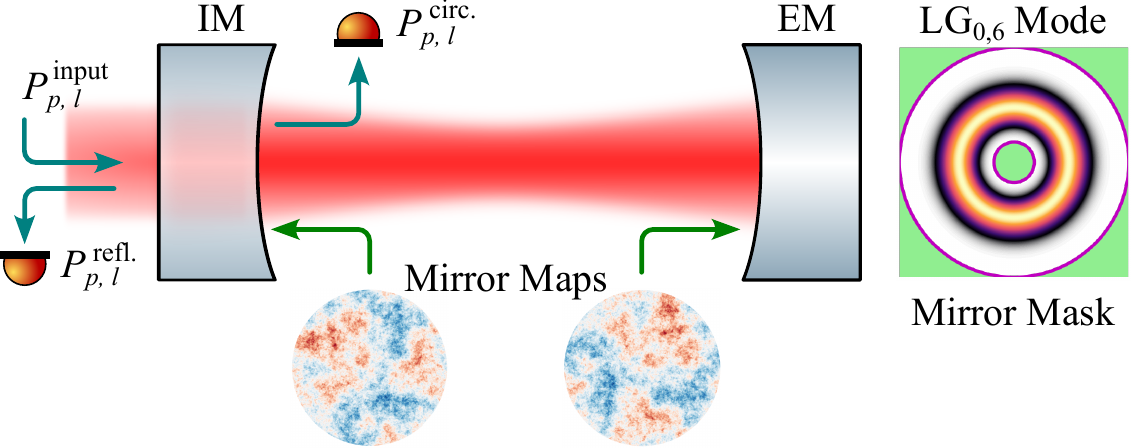}
    \caption{Schematic of a Virgo-like symmetric cavity used in the \texttt{Finesse} simulation. Two distinct random synthetic mirror maps are applied to the input and end mirrors. A circular mask (green indicates the AR-coated region or areas outside the mirror) is applied to the end mirror, with negligible overlap between the \LGsix{} mode and the AR-coated region.}
    \label{fig-FP_cav}
\end{figure}

We evaluate the performance of the proposed \LGsix{} mode under realistic mirror surface distortions in typical gravitational-wave interferometer working conditions. For this purpose, we consider a Virgo-like optical cavity, as illustrated in Fig.~\ref{fig-FP_cav}, with a cavity length of $2998~\mathrm{m}$. For simplicity, we assume a symmetric cavity in which both the input mirror (IM) and the end mirror (EM) have the same radius of curvature. IM has a power transmission of 0.014, and EM has a small transmission of $5~\mathrm{ppm}$. The nominal cavity finesse value under no additional mirror surface imperfection is around 445. The \texttt{Finesse} model includes spatial modes up to order 16, which was found to be sufficient for studying surface-error-induced mode scattering and losses of the sixth-order \LGsix{} mode based on convergence checks. The beam radii are rescaled to ensure that all modes exhibit the same clipping loss, due to their distinct intensity distributions. The resulting beam radii and corresponding mirror radii of curvature are summarized in Tab.~\ref{tab-roc_homs}, with the clipping loss on each test mass maintained at $1~\mathrm{ppm}$. The corresponding cavity $g$-factor products, $g_1 g_2$, are approximately $0.94$ and $0.77$ for the symmetric \LGzero{} and \LGsix{} configurations, respectively. Both configurations therefore remain geometrically stable, satisfying $0 < g_1 g_2 < 1$.

We assume that both test masses are perturbed from ideal spherical surfaces by random mirror maps. Under these imperfections, we compute the mode loss, mode power gain, and contrast defect using two independently simulated cavities representing the two interferometer arms. Since the arm cavities are over-coupled, we extract the mode loss $\Gamma_{p,\ell}$ for an injected \LGpl{} mode from the reflected field while the cavity is held on resonance for that mode:
\begin{equation}
    \Gamma_{p,\ell} = 1 - \frac{P_{p,\ell}^{\mathrm{refl.}}}{\widetilde{P_{p,\ell}^{\mathrm{refl.}}}} \, ,
\label{eq-loss}
\end{equation}
where quantities with a tilde denote the case without mirror surface maps, and those without denote the perturbed case. $\Gamma_{p,\ell}$ quantifies the fraction of mode power lost due to the presence of the mirror imperfections. This loss directly impacts, for example, the efficiency of squeezed light injected from the antisymmetric port.

To quantify the cavity power in the presence of mirror surface errors, we define the mode power gain $\mathcal{G}_{p,\ell}$ as
\begin{equation}
    \mathcal{G}_{p,\ell} = \frac{P_{p,\ell}^{\mathrm{circ.}}}{P_{p,\ell}^{\mathrm{input}}} \, ,
\label{eq-gain}
\end{equation}
which represents the ratio of circulating power in the inject mode inside the cavity to the input power. A higher $\mathcal{G}_{p,\ell}$ indicates more efficient power buildup for the inject, signal-carrying mode, reducing the required input power to achieve the desired arm power levels at hundreds of kilowatts to megawatts in gravitational-wave detectors to reach the quantum shot-noise targets.

\begin{figure*}[t]
    \centering
    \includegraphics[width=\linewidth]{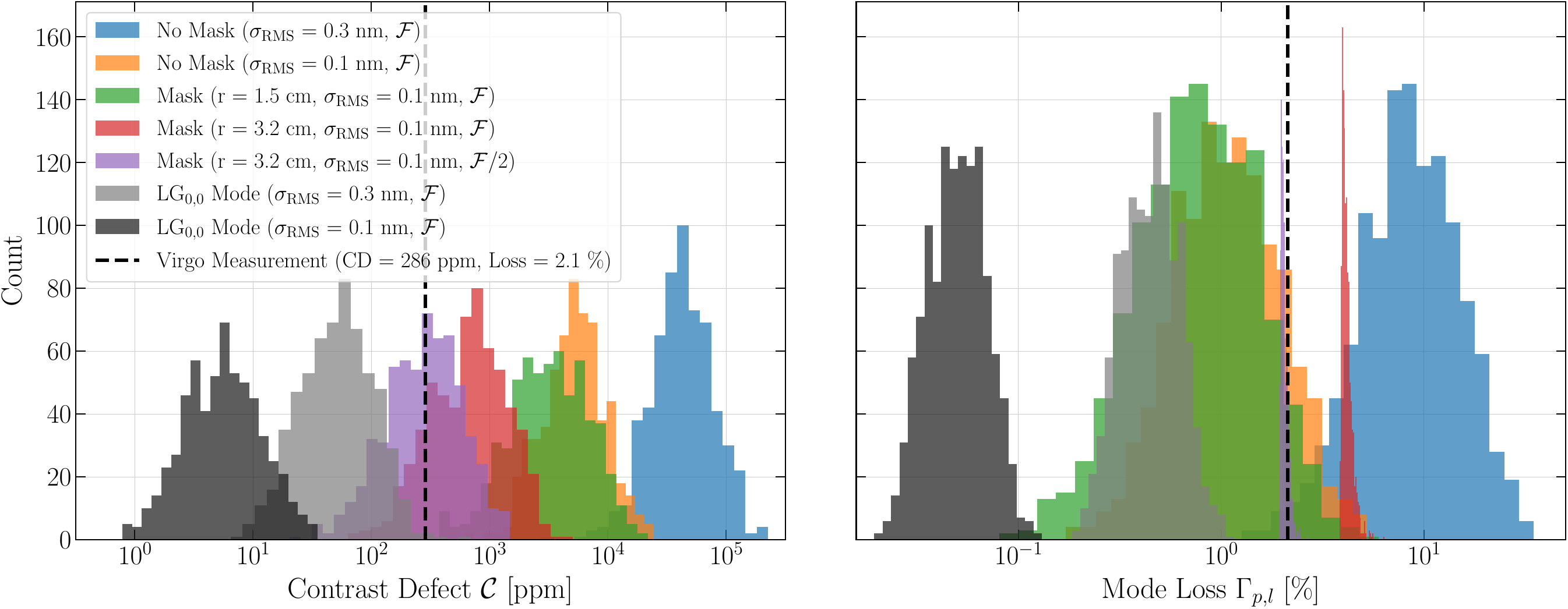}
    \caption{Histogram distributions of the contrast defect (left) and the mode loss (right) from the Monte Carlo simulations. The measured contrast defect and arm cavity mode loss in Virgo are indicated by the vertical dashed lines.}
    \label{fig-hist_CD_loss}
\end{figure*}

\begin{table*}[htbp]
\centering
\caption{Summary of statistics for contrast defect (ppm) and cavity mode loss (\%) across various cases, including increasing circular mask size, reduced mirror RMS error, and lowered cavity finesse. The assumed Virgo-like cavity has a nominal finesse $\mathcal{F}\sim445$.}
\setlength{\tabcolsep}{6pt}
\begin{tabular}{c|cccc|cccc}
\hline
\hline
\multirow{2}{*}{\diagbox{Case}{Quantity}} & \multicolumn{4}{c|}{Contrast Defect (ppm)} & \multicolumn{4}{c}{Mode Loss (\%)} \\
\cline{2-9}
 & Min. & Max. & Avg. & STD & Min. & Max. & Avg. & STD \\
\hline
No Mask ($\sigma_{\mathrm{RMS}}$=0.3~nm, $\mathcal{F}$) & 2685 & 226810 & 49994 & 31016 & 1.26 & 34.64 & 9.82 & 5.54 \\ 
No Mask ($\sigma_{\mathrm{RMS}}$=0.1~nm, $\mathcal{F}$) & 618 & 24502 & 6108 & 3808 & 0.17 & 5.25 & 1.28 & 0.80 \\ 
Mask (r=1.5~cm, $\sigma_{\mathrm{RMS}}$=0.1~nm, $\mathcal{F}$) & 374 & 21691 & 4145 & 3150 & 0.08 & 5.96 & 0.94 & 0.65 \\ 
Mask (r=3.2~cm, $\sigma_{\mathrm{RMS}}$=0.1~nm, $\mathcal{F}$) & 63 & 5056 & 847 & 609 & 3.83 & 6.39 & 4.17 & 0.26 \\ 
Mask (r=3.2~cm, $\sigma_{\mathrm{RMS}}$=0.1~nm, $\mathcal{F}$/2) & 20 & 1476 & 352 & 253 & 1.92 & 2.69 & 2.05 & 0.09 \\ 
\LGzero{} ($\sigma_{\mathrm{RMS}}$=0.3~nm, $\mathcal{F}$) & 6 & 692 & 65 & 53 & 0.18 & 1.16 & 0.47 & 0.16 \\ 
\LGzero{} ($\sigma_{\mathrm{RMS}}$=0.1~nm, $\mathcal{F}$) & 1 & 35 & 7 & 6 & 0.02 & 0.13 & 0.05 & 0.02 \\ 
\hline
Virgo Measurement & \multicolumn{4}{c|}{286 $\pm$ 24} & \multicolumn{4}{c}{2.12} \\
\hline
\hline
\end{tabular}
\label{tab-mean_std_combined}
\end{table*}

Finally, we use the contrast defect to quantify the beam quality and the excess light at the dark port of a Michelson interferometer. It is defined as the ratio of the optical power at the dark port to the total circulating power in the main interferometer. For an interferometer operating on the dark fringe, it is given by~\cite{Bond2017}
\begin{equation}
\mathcal{C} = \frac{\int_A \left|E_x - E_y\right|^2 \, dA}{\int_A \left|E_x + E_y\right|^2 \, dA} \, ,
\label{eq-contrast}
\end{equation}
where $E_x$ and $E_y$ correspond to the reflected fields from the two independently simulated cavities with random mirror perturbations, representing the fields returning from the two arms to the central beamsplitter. Imbalanced higher-order mode scattering in the two interferometer arms leads to increased light at the output port, thereby increasing the contrast defect, degrading interferometric dark-fringe quality, and worsening the overall strain sensitivity, and therefore must be mitigated. We note that Eqs.~(\ref{eq-loss}) and~(\ref{eq-gain}) refer specifically to the loss and gain of the targeted mode shape, while Eq.~(\ref{eq-contrast}) is computed from the full optical field. The contrast defect therefore includes the effect of all simulated higher-order-mode content and, in particular, the modal imbalance between the two arms.

To obtain statistically robust results, we perform a Monte Carlo analysis over many randomized trials. We perform 600 pairs of cavity simulations, each using two independently generated mirror maps for the input and output mirrors. For each configuration, we therefore obtain 600 contrast defect values and 1200 samples of mode loss and mode power gain. Statistical quantities such as the mean and standard deviation are observed to converge for this number of trials. 

\begin{figure*}[htbp]
    \centering
    \includegraphics[width=\linewidth]{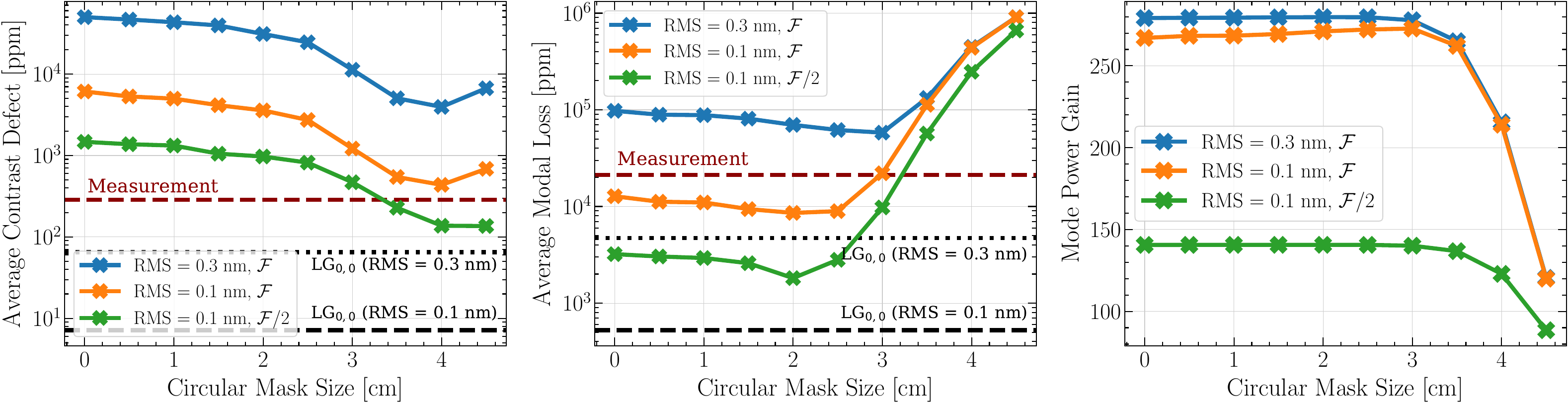}
    \caption{Average contrast defect (left), mode loss (middle), and mode power gain (right) as functions of the circular mask size. The currently measured values in Virgo are shown as horizontal red lines. The values for the fundamental Gaussian mode (\LGzero{}) are indicated by black dotted and dashed lines for RMS of 0.3~nm and 0.1~nm, respectively, for reference.}
    \label{fig-average_CD_loss_AG_r_mask}
\end{figure*}

\subsection{Monte Carlo results}
To evaluate the incremental benefits of a circular mirror mask with varying radius, improved mirror surface quality (from $0.3~\mathrm{nm}$ to $0.1~\mathrm{nm}$ RMS), and reduced cavity finesse, we analyze the following cases:

\begin{enumerate}
    \item Circular mask radii ranging from $0$ (no mask) to $4.5~\mathrm{cm}$ in $0.5~\mathrm{cm}$ increments,
    \item Two mirror surface RMS values: $0.3~\mathrm{nm}$ (current state-of-the-art) and $0.1~\mathrm{nm}$ (anticipated near-future performance),
    \item A reduced cavity finesse case by half for the $0.1~\mathrm{nm}$ RMS scenario,
    \item Equivalent simulations for the fundamental \LGzero{} mode as a reference study.
\end{enumerate}

For each configuration, statistical quantities such as the mean and standard deviation from the Monte Carlo simulations are extracted. Fig.~\ref{fig-hist_CD_loss} shows the histogram distributions of the contrast defect and mode loss for various representative cases, including no mask, different RMS values, varying mask sizes, reduced finesse, and the \LGzero{} mode under different surface conditions.

In addition, we indicate typical measured values from Virgo for comparison. The contrast defect during O4 is approximately $286 \pm 24~\mathrm{ppm}$, and a representative arm cavity round-trip loss is approximately $75~\mathrm{ppm}$~\cite{VirgoCollaboration:25, virgo_logbook_68368}. This corresponds to a measured mode loss
\begin{equation}
    \Gamma_{\mathrm{mea.}}\sim \frac{2 \mathcal{F}}{\pi} \times 75~\mathrm{ppm} \approx 2.12\% \, ,
\end{equation}
which are shown in Fig.~\ref{fig-hist_CD_loss} are the vertical dashed lines. The corresponding statistical values for all cases are summarized in Tab.~\ref{tab-mean_std_combined}. The contrast defect and mode loss decrease as the circular mask is introduced, alongside improvements in mirror surface quality and a reduction in cavity finesse. It should be noted that, in realistic interferometer operation, additional imperfections beyond mirror-surface distortions can further degrade the overall contrast defect. For example, differential mode-matching imbalance between the two arm cavities provides an additional contribution to the contrast defect~\cite{VirgoCollaboration:25}. These effects are not included in the present \LGzero{} simulation, and may therefore account for the remaining discrepancy between the simulated contrast defect and the currently measured values in Virgo.

Fig.~\ref{fig-average_CD_loss_AG_r_mask} shows the averaged contrast defect, mode loss, and mode power gain as functions of the circular mask radius for both RMS values and the reduced-finesse case. Each data point is extracted from 600 pairs of Monte Carlo cavity simulations. The contrast defect decreases significantly as a circular mask is introduced and its radius increases. An optimal mask radius of $4~\mathrm{cm}$ yields approximately an order-of-magnitude reduction in contrast defect. Increasing the mask size beyond this point leads to additional scattering losses from the mask itself, causing the contrast defect to increase again.

Improving the mirror surface quality from $0.3~\mathrm{nm}$ to $0.1~\mathrm{nm}$ RMS provides an additional order-of-magnitude reduction in contrast defect. Furthermore, reducing the cavity finesse by a factor of two yields an additional reduction by approximately a factor of $5$. Combining these mitigation strategies, including the use of optimal circular masking, improved surface quality, and reduced finesse, results in an average contrast defect for \LGsix{} of only $140~\mathrm{ppm}$, approximately a factor of two below typical Virgo measurements.

On the other hand, a large circular mask introduces additional mode loss for the \LGsix{} mode, thereby reducing the cavity power gain. To maintain the total mode loss below the measured value of $2.12\%$, the maximum allowable mask radius is $3.2~\mathrm{cm}$ in the low-RMS, reduced-finesse configuration. This defines the optimal mask size based on the trade-off between minimizing contrast defect and limiting excess loss.

From the mode power gain shown in the rightmost panel of Fig.~\ref{fig-average_CD_loss_AG_r_mask}, at a mask radius of $3.2~\mathrm{cm}$, the \LGsix{} power buildup has not yet begun to degrade significantly, indicating that no additional input power requirement is imposed on the pre-stabilized laser system to reach the same target interferometer power. At this configuration, the average contrast defect is also consistent with current measurement values, further demonstrating its feasibility.

\begin{table}[t]
\centering
\caption{Single-bounce power loss and corresponding OMC power transmission for the degenerate \LGpl{} modes of the same order as the desired \LGsix{} mode, for a circular mask with radius $0.7\,w$ (3.2~cm for a beam radius of 4.6~cm) applied to an OMC end mirror.}
\setlength{\tabcolsep}{6pt}
\begin{tabular}{c|c|c|c|c}
\hline 
\hline 
Degenerate Mode & $\mathrm{LG}_{0, 6}$ & $\mathrm{LG}_{1, 4}$ & $\mathrm{LG}_{2, 2}$ & $\mathrm{LG}_{3, 0}$ \\
\hline  
Single Bounce [\%] & $7\cdot 10^{-3}$ & 1 &  16 & 16 \\ \hline 
Throughput [\%] & 97.6 & 11.2 & 0.11 & 0.11  \\ \hline \hline 
\end{tabular}
\label{tab-OMC_filter}
\end{table}

In addition, in gravitational-wave detectors operating with the fundamental mode, higher-order spatial modes scattered by mirror surface imperfections and leaking to the dark port can be effectively suppressed by the bowtie Output Mode Cleaner (OMC). The OMC provides additional spatial filtering, thereby further reducing the contrast defect. For instance, in Virgo operation, the OMC further suppresses the contrast defect by approximately a factor of $2 \cdot 10^{-4}$, reducing it from $(2.86 \pm 0.24)\cdot 10^{-4}$ before the OMC to $6 \cdot 10^{-8}$ after the OMC~\cite{VirgoCollaboration:25}. 

However, this approach is generally ineffective when higher-order modes are used. In such cases, degenerate modes share the same resonance condition within the OMC cavity and are therefore transmitted alongside the desired signal-carrying mode. As a result, unwanted mode content reaches the ultimate photodetector, leading to a degradation in the signal-to-noise ratio.

On the other hand, for operation with the \LGsix{} mode, a similar mitigation strategy can be employed for the output optics. A circular mask with an AR-coated central region can be introduced on one of the OMC end mirrors. As illustrated in Fig.~\ref{fig-power_loss_LGpl_r_mask}, this mask induces selective optical losses for the unwanted degenerate LG modes while largely preserving the desired \LGsix{} mode. For instance, Tab.~\ref{tab-OMC_filter} presents the single-bounce power loss and the corresponding cumulative OMC transmission for a circular mask radius of $0.7\,w$, which corresponds to the optimal mask size of 3.2~cm for a beam radius of 4.6~cm identified for the arm cavity. The results show that only $11.2\%$ of the $\mathrm{LG}_{1,4}$ mode power survives after transmission through the OMC, while the $\mathrm{LG}_{2,2}$ and $\mathrm{LG}_{3,0}$ modes are suppressed to $0.11\%$ of their initial power. This demonstrates that the introduction of a tailored circular mask to the output optics can strongly suppress unwanted degenerate modes while preserving high transmission for the signal-carrying $\mathrm{LG}_{0,6}$ mode. 

In addition to the output optics, the implementation of the \LGsix{} mode also places requirements on the input laser and squeezed-light injection systems. For instance, the pre-stabilized laser and mode-generation system must generate and deliver the required mode shape with sufficiently high purity~\cite{Mueller_2016}. Although the target arm power does not introduce a new input-power requirement, the supplied laser power may need to increase depending on the efficiency of the mode-generation method~\cite{10.1063/5.0137085}.
 
\section{Conclusion and discussion \label{sec-conclusion}}
Previously, higher-order Laguerre-Gaussian modes, such as the \LGthree{} and \LGtwo{} modes, have been investigated as promising alternatives to reduce test-mass thermal noise in laser interferometric gravitational-wave detectors. However, these modes have been shown to exhibit extreme sensitivity to mirror surface deformations, which has so far limited their practical applicability in gravitational-wave detectors. In this work, we investigate an alternative ``donut-shaped'' \LGol{} mode as a replacement for the currently employed fundamental Gaussian beam to achieve significant thermal noise reduction. We examine the robustness of the \LGsix{} mode against realistic mirror surface figure errors present in current gravitational-wave detectors and propose mitigation strategies to limit the resulting beam quality degradation. Three primary strategies are explored: (i) utilizing an optimally-designed circular mirror mask to selectively increase losses for degenerate modes, (ii) improving mirror surface error to anticipated near-future levels, and (iii) reducing the arm cavity finesse. Monte Carlo simulations over many randomized trials demonstrate that, by combining circular masking with engineered selective losses for degenerate modes, improved mirror surface quality, and reduced cavity finesse, the overall contrast defect can be reduced by more than two orders of magnitude and the mode loss decreased by nearly a factor of five, allowing the injected \LGsix{} mode to recover beam quality and optical performance at levels comparable to those observed in current gravitational-wave detectors. A practical caveat of the proposed circular coating profile is that the outer HR-coated annulus must maintain sufficiently low surface roughness and coating-thickness inhomogeneity. While existing coating measurements show some degradation of these quantities away from the central region~\cite{Pinard:17, Degallaix:19}, this is likely a consequence of coating R\&D historically optimized for fundamental Gaussian beams rather than an intrinsic limitation, motivating dedicated deposition studies for annular, higher-order mode illumination. In addition, the impact of relative beam-mask offset was also evaluated, showing that the contrast defect is less sensitive to off-centering than the loss, and that keeping the relative offset below $\sim 2~\mathrm{mm}$ limits the loss increase to below $\sim 10\%$. 

This work opens up new research and development pathways for employing \LGol{}-type modes that achieve significant thermal noise reduction while maintaining beam quality and optical performance comparable to current gravitational-wave interferometers. Future work includes experimental validation of the \LGsix{} mode in tabletop and prototype interferometers, investigation of the impact of relative beam-centering with respect to the circular mask, and more detailed studies of alignment-control requirements, point-defect robustness, mode-matching tolerances, and compatibility with squeezed-light injection and other advanced interferometer configurations. In addition, extending the present analysis from simplified cavity studies to full interferometric gravitational-wave detector configurations will be essential to comprehensively evaluate the performance, robustness, and control implications of higher-order LG modes in realistic detector environments.

\begin{acknowledgments}
The authors thank Anna Green for helpful comments during the preparation of this manuscript. The authors acknowledge the use of the ITM04 mirror surface map data from LIGO. The authors acknowledge support from ANR-18-IDEX-0001 and ANR-23-CE31-0004. This document was submitted to the Virgo and LIGO collaborations under the document numbers VIR-0404A-26 and P2600354, respectively.

\end{acknowledgments}

\bibliography{references}

\end{document}